\documentclass[a4paper,english,cleveref, autoref, thm-restate, pdfa]{lipics-v2021}
\pdfoutput=1 %uncomment to ensure pdflatex processing (mandatatory e.g. to submit to arXiv)
\hideLIPIcs  %uncomment to remove references to LIPIcs series (logo, DOI, ...), e.g. when preparing a pre-final version to be uploaded to arXiv or another public repository
\nolinenumbers
\usepackage{amssymb}
\usepackage{scalerel}
\usepackage{todonotes}
\usepackage{graphicx}

\title{Tempus fugit: Anyone can understand temporal logic if they have to save the realm}

\titlerunning{Tempus fugit: Anyone can understand temporal logic}

\author{Benjamin Bisping}{Télécom SudParis, Institut Polytechnique de Paris, France \and \url{https://bbisping.de/}}{benjamin.bisping@telecom-sudparis.eu}{https://orcid.org/0000-0002-0637-0171}{}
\author{Tobias Loch}{Technische Universität Berlin, Germany}{tobias.loch@mail.de}{}{}
\author{Mustafa Mohsen}{Technische Universität Berlin, Germany}{mohsen@tu-berlin.de}{}{}
\author{Alessio Nicolo {Perna}}{Technische Universität Berlin, Germany}{alessio.perna@hotmail.de}{}{}
\author{Maximilian Lukas {Stamm}}{Technische Universität Berlin, Germany}{}{}{}

\authorrunning{B. Bisping, T. Loch, M. Mohsen, A. N. Perna, M. L. Stamm}

\Copyright{Benjamin Bisping, Tobias Loch, Mustafa Mohsen, Alessio Nicolo Perna, Maximilian Lukas Stamm}



\category{} %optional, e.g. invited paper

\relatedversion{} %optional, e.g. full version hosted on arXiv, HAL, or other respository/website
\funding{Benjamin Bisping is partially funded by the Orange and Télécom SudParis CRE Project “Formal validation of the singularization principle.”}%optional, to capture a funding statement, which applies to all authors. Please enter author specific funding statements as fifth argument of the \author macro.

\acknowledgements{
  Apart from the paper authors, Florian Eyert and Kejni Dema contributed to the development of \emph{Tempus fugit}.
  We thank Philipp Schlehuber-Caissier and Arno Wilhelm-Weidner for feedback on a draft version of the paper.
}

\newcommand{\pltl}{\ensuremath{\mathsf{PLTL}}}

\newcommand{\propPart}[1]{\textcolor{red!60!black}{#1}}
\newcommand{\futurePart}[1]{\textcolor{blue!55!black}{#1}}
\newcommand{\pastPart}[1]{\textcolor{green!50!black}{#1}}
\newcommand{\pastOverlay}[2][0.27ex]{\ooalign{$#2$\cr\hidewidth\raisebox{#1}{$\scriptscriptstyle\circlearrowleft$}\hidewidth\cr}}
\newcommand{\opNext}{\futurePart{\bigcirc}}
\newcommand{\opGlobally}{\futurePart{\Box}}
\newcommand{\opEventually}{\futurePart{\rotatebox[origin=c]{45}{$\Box$}}}
\newcommand{\opUntil}{\futurePart{\mathbf{U}}}
\newcommand{\opPrevious}{\pastPart{\pastOverlay{\bigcirc}}}
\newcommand{\opHistorically}{\pastPart{\pastOverlay[0.45ex]{\Box}}}
\newcommand{\opOnce}{\pastPart{\pastOverlay[0.5ex]{\rotatebox[origin=c]{45}{$\Box$}}}}
\newcommand{\opSince}{\pastPart{\mathbf{S}}}

\begin{document}

\maketitle

\begin{abstract}
Often, the easiest way to learn something is to have to use it for a purpose.
This purpose can be playful:
In \emph{Tempus fugit}, the player takes on the role of a magician who has to defeat enemies by casting spells.
The applicability of spells and enemy attacks depends on the truth of formulas in linear temporal logic with past
with respect to a trace that the player gradually builds.
So, whoever wants to save the realm from monsters has to learn to read logic formulas.

This paper describes the small browser game and explains our design choices.
We expose how game mechanics connect to linear temporal logic with past over finite traces,
and how this can help players approach a daunting topic like formal logic.
\end{abstract}

\section{Introduction}
\label{sec:intro}

%There is no official statistics on logics literacy, is there? -> maybe https://www.frontiersin.org/journals/education/articles/10.3389/feduc.2023.1247653/full
Sadly, most people never learn to read formal logics, let alone temporal and modal ones.
%As shown by Fehér et al. \cite{Feher2023}, even in universities, merely propositional logic tasks were at best answered correctly 67\% of the time.% Strange source; also, the paper's claim is not really about making people understand logic better than a course, but about growing the audience of logic courses.
Despite persistent preaching on the merits of such logics, for instance by Lamport~\cite{lamport1983whatGoodTL},
and its inclusion in many curricula,
the majority of the programmer population lives under the assumption that temporal logics are an arcane art, something to be discarded as too complicated.
But the core concepts are easy to grasp---even in the context of a science communication event or a chat at a family gathering.
How do we help people take the first step beyond the initial barriers of notation and limiting beliefs?

\emph{Tempus fugit} is a small browser game in which one plays a magician.
Which spells one can cast depends on atomic propositions that players set over time.
The twist: Spell cards express their applicability conditions in \emph{linear temporal logic with past} ($\pltl$), as do enemy abilities.
In order to play effectively, players must thus learn to read formulas of $\pltl$.

By \emph{framing the challenge in terms of magic in a computer game}, we “trick” players into engaging with temporal logic.
Implicitly, strange logical notation becomes part of the fun.
Limiting beliefs like “Formal logic is made for people who are more mathematical than me” are overwritten by “This game is made to be played by people like me.”

\subparagraph{Contribution}
In this paper, we present \emph{Tempus fugit}, a computer game that can be employed to convey the basics of temporal logic to everyday people, for instance, in science communication contexts or at the beginning of a logic course.

Players of the game implicitly learn how to manipulate valuations of atomic propositions in order to make propositional and temporal formulas true or false.
This seeds the concepts of atomic propositions, valuations, and traces, as well as operators, formulas, and truth-value semantics.

\subparagraph{Structure}
This paper recalls definitions of temporal logic $\pltl$ (\autoref{sec:theory}),
describes the game mechanics and how they relate to $\pltl$ (\autoref{sec:game})
and discusses the choices we have made in design and implementation (\autoref{sec:design}).
\autoref{sec:conclusion} closes with some remarks on limits and opportunities in the context of “games as learning machines.”

\subparagraph{Availability}
\emph{Tempus fugit} can be played directly in the browser on \url{https://benkeks.itch.io/tempusfugit}.
Its TypeScript source code is available freely on \url{https://github.com/benkeks/tempus-fugit}.

\subparagraph{Demo}
The game will be demoed at the workshop \href{https://teal.cs.brown.edu/floc2026/}{\emph{TEAL 2026: Tools for Educational Activities in Logic}} during \emph{FLoC 2026} in Lisbon.
If you cannot make it there, we are happy to repeat it elsewhere.
%A demo of the game would show a few minutes of game play, and relate it to the basic concepts of linear temporal logic.
Because the game runs in the browser and does not require accounts, participants can easily be invited to try their luck in playing the game on their own devices.

\section{Background: Linear temporal logic}
\label{sec:theory}

This section goes over the syntax and semantics of the formal logic relevant to \textit{Tempus fugit}.
Linear temporal logic~(LTL) has been introduced for program verification by Pnueli~\cite{Pnueli1977}.
It extends classic propositional logic with modalities to describe how variables change over time.
We use a variant that also contains past-modalities~\cite{gabbay1987pastTL}.
Its syntax can be summarized as follows:

\begin{definition}[Syntax of \pltl]
\label{def:pltl-syntax}
Let $AP$ be a set of atomic propositions.
Formulas $\varphi$ of \pltl\ are generated by the grammar
\[
\begin{array}{r@{~}c@{~}l@{\qquad}l}
\varphi
  & ::= & \propPart{p}
          \mid \propPart{\neg}\varphi
          \mid \varphi\,\propPart{\wedge}\,\varphi
          \mid \varphi\,\propPart{\vee}\,\varphi
          \mid \varphi\,\propPart{\leftrightarrow}\,\varphi
  & \propPart{\textit{(propositional: atom, neg, conj, disj, iff)}} \\[4pt]
    &  &    \mid \opNext\,\varphi
      \mid \opGlobally\,\varphi
      \mid \opEventually\,\varphi
      \mid \varphi\,\opUntil\,\varphi
  & \futurePart{\textit{(future: next, globally, eventually, until)}} \\[4pt]
    & &     \mid \opPrevious\,\varphi
      \mid \opHistorically\,\varphi
      \mid \opOnce\,\varphi
      \mid \varphi\,\opSince\,\varphi
  & \pastPart{\textit{(past: previous, historically, once, since)}}
\end{array}
\]
where $p \in AP$.
\end{definition}

LTL can be interpreted over either infinite or finite sequences of states, called traces.
While the original formulation assumes infinite traces, finite trace semantics have been studied and are often assumed or used in real applications, see~\cite{dgv2023ltlFin}.

\begin{definition}[Finite-trace semantics of \pltl]
\label{def:pltl-semantics}
A \emph{finite trace} is a non-empty sequence $\sigma = \sigma_0\,\sigma_1\cdots\sigma_{n-1}$
with each $\sigma_i \subseteq AP$.
Satisfaction $\sigma, i \models \varphi$ at position $0 \le i < n$ is defined inductively:
\[
\renewcommand{\arraystretch}{1.0}
\begin{array}{@{}r@{~}c@{~}l@{~~}c@{~~}l@{}}
\multicolumn{5}{@{}l}{\propPart{\textit{Propositional:}}} \\
\sigma, i & \models & \propPart{p}
  & \Leftrightarrow & p \in \sigma_i \\
\sigma, i & \models & \propPart{\neg}\varphi
  & \Leftrightarrow & \sigma, i \not\models \varphi \\
\sigma, i & \models & \varphi\,\propPart{\land}\,\psi
  & \Leftrightarrow & \sigma, i \models \varphi \text{ and } \sigma, i \models \psi \\
\sigma, i & \models & \varphi\,\propPart{\lor}\,\psi
  & \Leftrightarrow & \sigma, i \models \varphi \text{ or } \sigma, i \models \psi \\
\sigma, i & \models & \varphi\,\propPart{\leftrightarrow}\,\psi
  & \Leftrightarrow & \sigma, i \models \varphi \text{ equals } \sigma, i \models \psi \\[4pt]
\multicolumn{5}{@{}l}{\futurePart{\textit{Future:}}} \\
\sigma, i & \models & \opNext\,\varphi
  & \Leftrightarrow & i{+}1 < n \text{ and } \sigma, i{+}1 \models \varphi \\
\sigma, i & \models & \opGlobally\,\varphi
  & \Leftrightarrow & \text{for all } j \in [i,n) \text{ holds that } \sigma, j \models \varphi \\
\sigma, i & \models & \opEventually\,\varphi
  & \Leftrightarrow & \text{there is } j \in [i,n)\text{ with } \sigma, j \models \varphi \\
\sigma, i & \models & \varphi\,\opUntil\,\psi
  & \Leftrightarrow & \text{there is } j \in [i,n)\text{ with } \sigma, j \models \psi \text{ and for all } k \in [i,j)\text{ holds that } \sigma, k \models \varphi \\[4pt]
\multicolumn{5}{@{}l}{\pastPart{\textit{Past:}}} \\
\sigma, i & \models & \opPrevious\,\varphi
  & \Leftrightarrow & i > 0 \text{ and } \sigma, i{-}1 \models \varphi \\
\sigma, i & \models & \opHistorically\,\varphi
  & \Leftrightarrow & \text{for all } j \in [0,i]\text{ holds that } \sigma, j \models \varphi \\
\sigma, i & \models & \opOnce\,\varphi
  & \Leftrightarrow & \text{there is } j \in [0,i]\text{ with } \sigma, j \models \varphi \\
\sigma, i & \models & \varphi\,\opSince\,\psi
  & \Leftrightarrow & \text{there is } j \in [0,i]\text{ with } \sigma, j \models \psi \text{ and for all } k \in (j,i]\text{ holds that } \sigma, k \models \varphi
\end{array}
\]
\end{definition}

\newcommand{\aplight}{\textcolor{yellow!80!black}{\texttt{L}}}
\newcommand{\aptransform}{\textcolor{violet!80!black}{\texttt{T}}}
\newcommand{\apnature}{\textcolor{blue!80!black}{\texttt{N}}}
\newcommand{\apstrength}{\textcolor{red!80!black}{\texttt{S}}}

\begin{example}
  \label{ex:pltl}
  Consider $\mathit{AP}_\mathsf{TF} := \{\aplight, \aptransform, \apnature, \apstrength \}$ and the formula $\varphi_\mathsf{L} := \opPrevious\opPrevious\apnature \land \opOnce (\aplight \land \aptransform) \land \apstrength$.
  Intuitively, it means that $\apnature$ was true two time steps ago, that $\aplight$ and $\aptransform$ were true at some point in the past, and that $\apstrength$ is true now.

  Consider the trace $\sigma_\mathsf{L} = \langle\{\aplight, \aptransform\}, \{\apnature\}, \{\apstrength\}, \{\apstrength\}\rangle$.
  Then $\sigma_\mathsf{L}, 3 \models \varphi_\mathsf{L}$, since $\sigma_\mathsf{L}, 1 \models \apnature$, $\sigma_\mathsf{L}, 0 \models \aplight \land \aptransform$, and $\sigma_\mathsf{L}, 3 \models \apstrength$.
  But $\sigma_\mathsf{L}, 2 \not\models \varphi_\mathsf{L}$, since $\sigma_\mathsf{L}, 2 \not\models \opPrevious\opPrevious\apnature$.
  Different traces might be models of the same formula.
  For instance,
  $\sigma_\mathsf{R} = \langle
  \{\apnature, \apstrength\},
  \{\aplight\},
  \{\aplight\},
  \{\apnature\},
  \{\aplight, \aptransform\},
  \{\aplight, \apstrength\},
  \varnothing,
  \varnothing
  \rangle$ also has
  $\sigma_\mathsf{R}, 5 \models \varphi_\mathsf{L}$.
\end{example}

\emph{Tempus fugit} operates over finite traces, with a trace growing incrementally as the player progresses through a level.
The atomic propositions consist of four runes manipulated by the player.
We disregard nuances like, for example, weak versions of the operators.

\section{Tempus fugit}
\label{sec:game}

From the perspective of mechanics, \emph{Tempus fugit} is a single-player turn-based deck-building game, where cards can be played according to a trace that is built over time.
Its pixel-art presentation draws from magic tropes.
In this section, we first describe the gameplay (\autoref{subsec:gameplay}), then relate it to $\pltl$ (\autoref{subsec:repr-ltl}), and finally explain the game narrative (\autoref{subsec:story}).

\begin{figure}
    \centering
    \includegraphics[width=\linewidth]{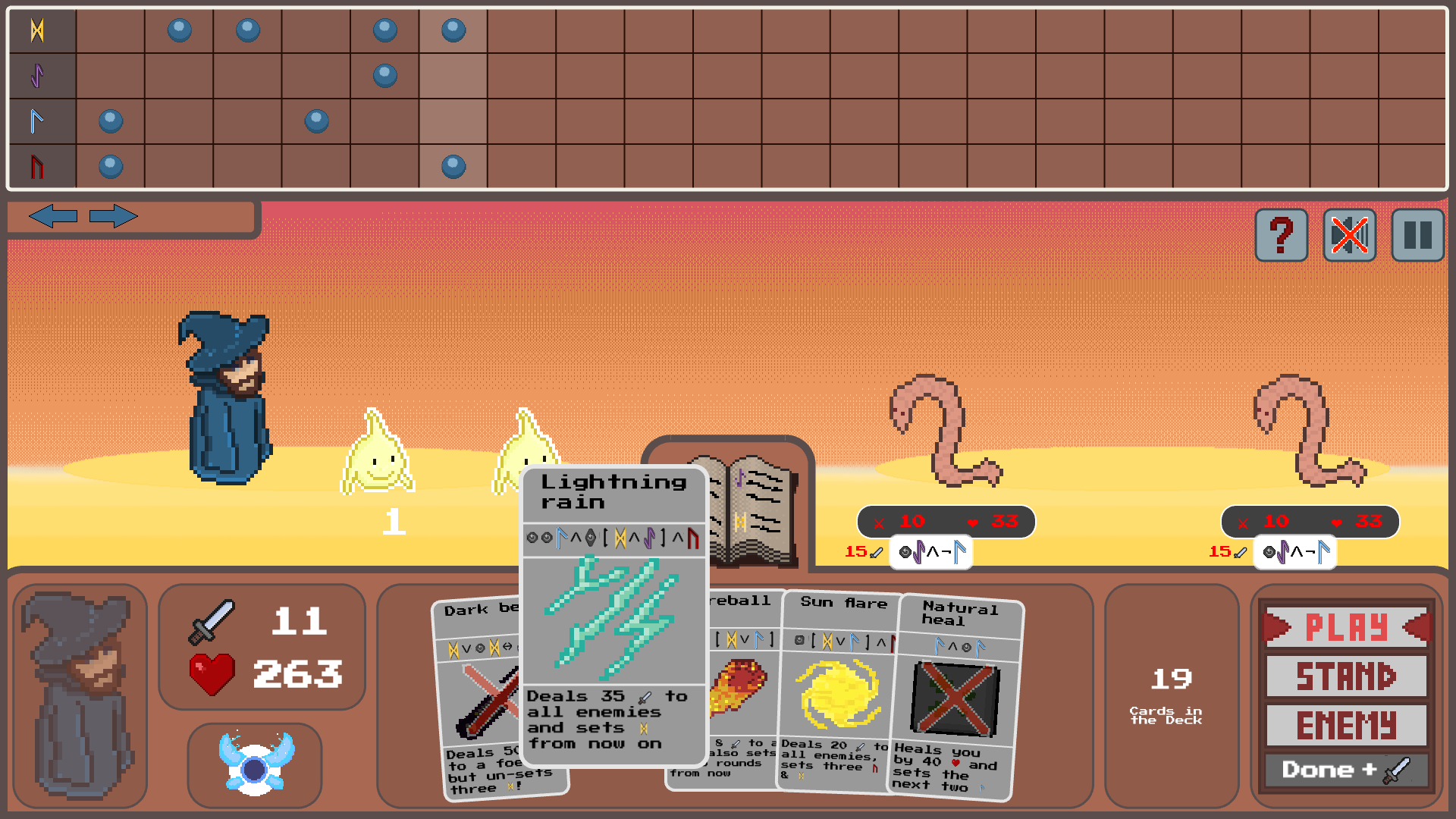}
    \caption{Main game scene of \emph{Tempus fugit} in level 6 “What worm is this?” with the player about to play a “Lightning rain” card.}
    \label{fig:screen-fight}
\end{figure}

\subsection{Mechanics and gameplay}
\label{subsec:gameplay}
%\emph{Tempus fugit} guides the user progression through a main map screen, presenting a linear path of levels. These levels are sequentially unlocked after completion of the previous one, with the first being unlocked by default. % redundant to later paragraph
A level of \emph{Tempus fugit} consists of a sequence of turn-based fights between the player and waves of enemies.
\autoref{fig:screen-fight} shows a typical fight scene.
Each fight is won by reducing the hitpoints of all enemies (right side of \autoref{fig:screen-fight}) to zero, and is lost if the player's (left side of \autoref{fig:screen-fight}) hitpoints drop to zero first.
Each enemy and the player have basic attack capabilities, which can be augmented by fulfilling logical conditions.
The player, in particular, can cast spells by playing cards from their hand (bottom panel of \autoref{fig:screen-fight}), which have $\pltl$ formulas as applicability conditions.
To cast spells, they have to manipulate a valuation of “runes” (atomic propositions in $\pltl$) over time (top panel in \autoref{fig:screen-fight}).

Each turn starts with the player drawing a card, and then discarding cards if they would have more than five cards in their hand.
They can then change the valuation of up to two runes at the current time index, and cast spells by dragging cards from their hand onto the spell book or onto affected enemies.
Most cards deal some damage to an enemy or heal the player.
Many have side effects that change the current or future valuation of runes.
Some of the cards summon so-called stands, which support the player for some turns if their conditions are still fulfilled in those turns.
After the player ends their turn, stands become active (if their conditions are met).
If the player has not cast any spell, they inflict some base damage onto the enemies and get an additional card in the next turn.
Then the enemies attack.
Each living enemy deals a base attack, and an additional attack if its $\pltl$ condition is satisfied.

The levels are connected by a small story told between fights.
Over time, the player gets access to more and more complex cards, and faces enemies with more and more complex conditions.
In between levels, the player can customize their card deck through a deck builder (\autoref{fig:deckbuilder-screen}) reachable from the map screen.

\subsection{Representation of logic and trace}
\label{subsec:repr-ltl}

\begin{figure}
    \centering
    \includegraphics[width=\linewidth]{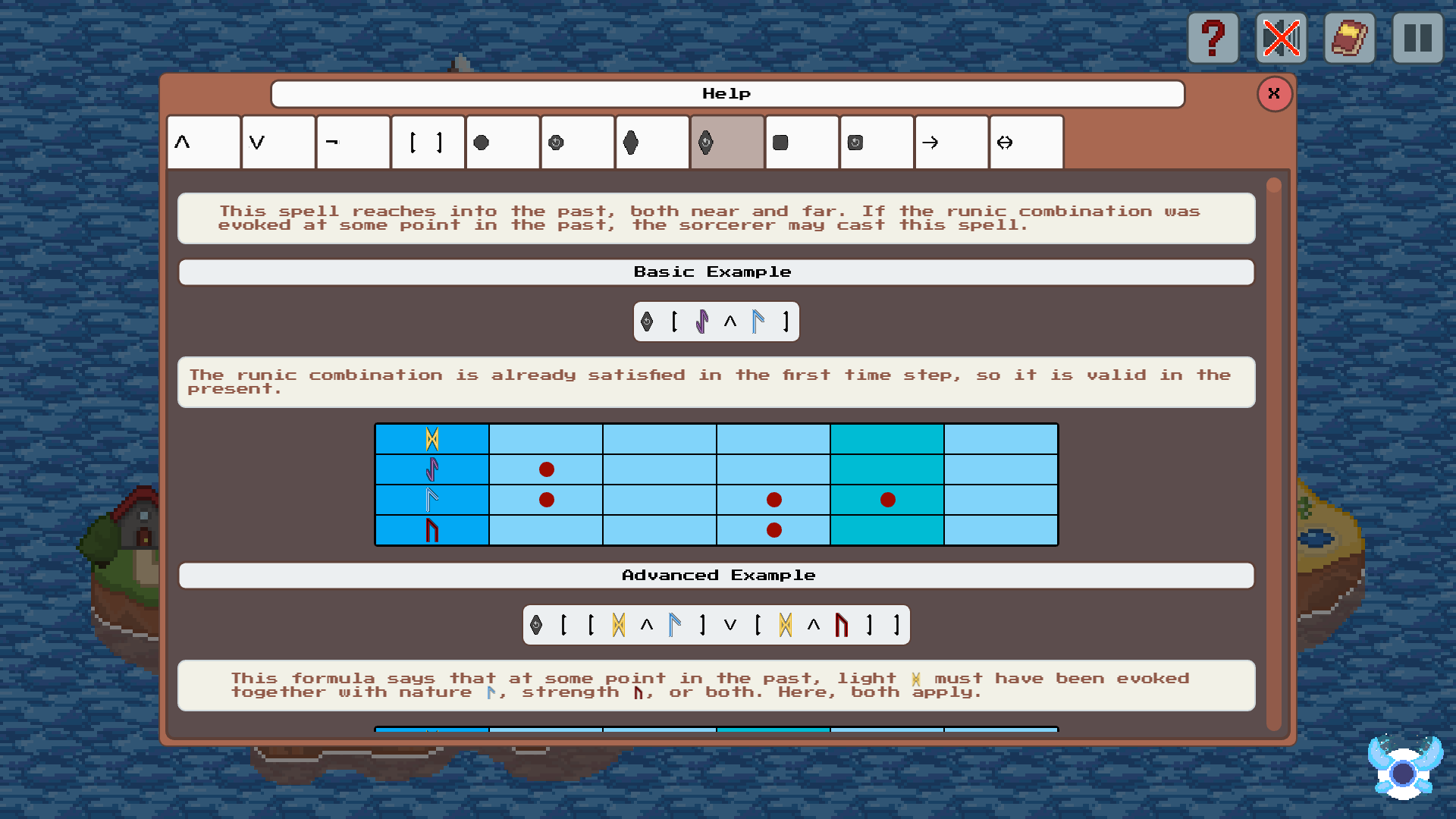}
    \caption{Help screen describing the \emph{once} modality.}
    \label{fig:help-screen}
\end{figure}

The core mechanic of \emph{Tempus fugit} is about making traces model certain temporal formulas.

$\pltl$ formulas feature on the cards and in enemy conditions of special attacks.
The logical connectives use almost-standard notation close to \autoref{def:pltl-syntax} with “undo” arrows to represent past modalities.
The atomic propositions $\mathit{AP}_\mathsf{TF}$ consist of four runes of different colors.
The formula on the “Lightning rain” card of \autoref{fig:screen-fight} matches exactly the example formula $\varphi_\mathsf{L}$ of \autoref{ex:pltl}.

The trace appears at the top of the screen with one column per time turn and one row per atomic proposition.
\autoref{fig:screen-fight} shows precisely trace $\sigma_\mathsf{R}$ at time step 5 of \autoref{ex:pltl}.
Traces are built incrementally with a “current time” progressing per turn ($i \in \mathbb{N}$ in the notation of \autoref{def:pltl-semantics}).
The player can change the valuation of two runes at the current time only, that is, not in the past or future.
However, some cards have side effects that change the valuation of runes in the past or future.
For example, “Lightning rain” of \autoref{fig:screen-fight} will activate $\aplight$ for the rest of the trace.

\begin{figure}
    \centering
    \includegraphics[width=\linewidth]{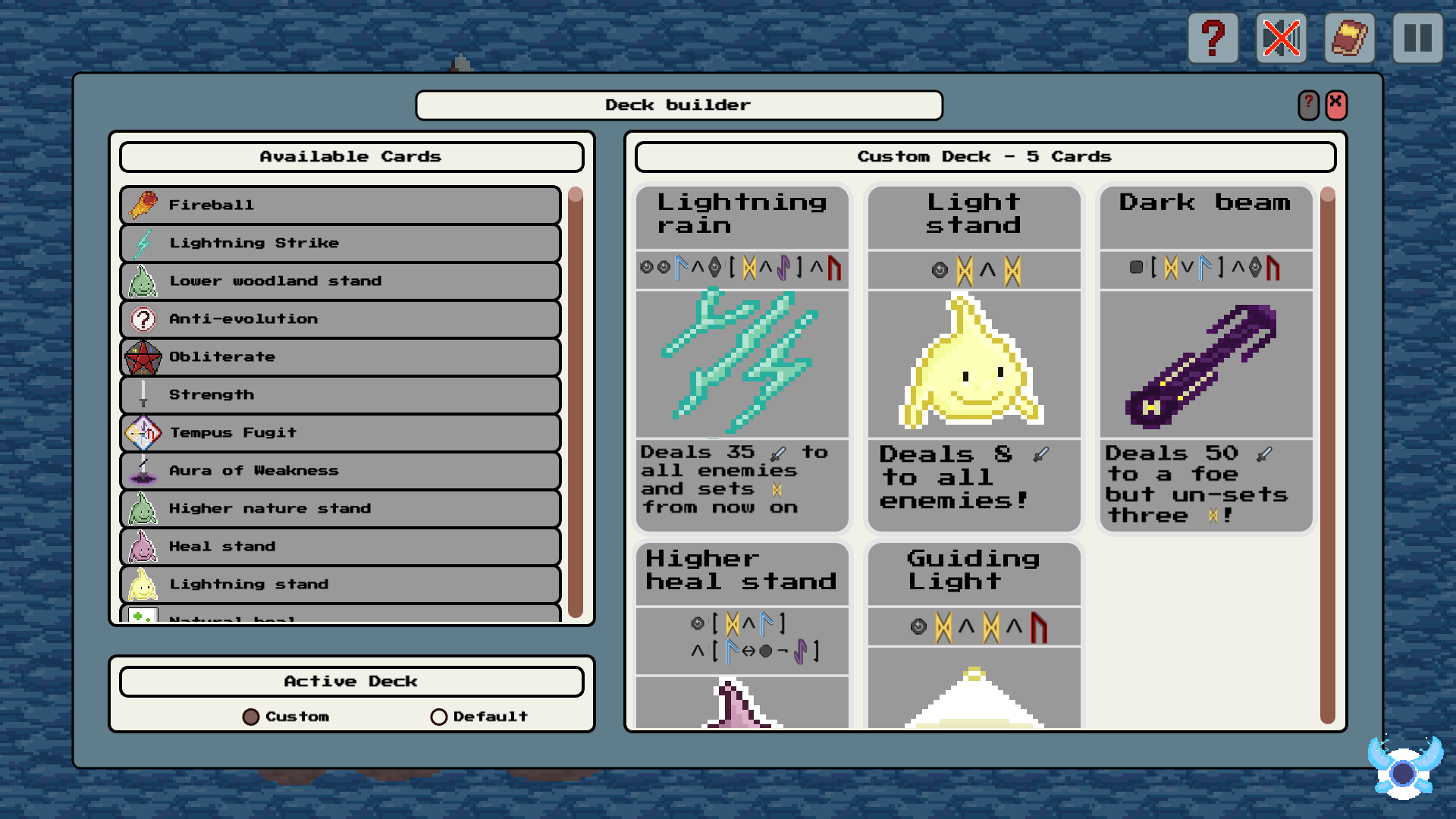}
    \caption{Deck builder combining “Lightning rain” with cards that benefit from global light activation.}
    \label{fig:deckbuilder-screen}
\end{figure}

Whether a card can be played depends on the truth value of its formula at the very time.
Therefore, the player should strategically manipulate the trace in order to satisfy the conditions of cards that they want to play and to falsify the enemy conditions.
\autoref{fig:deckbuilder-screen} shows the deck builder with cards that interact positively with “Lightning rain:”
For instance, it can always be followed up by “Dark beam” that has as activation condition $\opGlobally(\aplight \lor \apnature) \land \opOnce \apstrength$, which means that each future point must be filled with $\aplight$ or $\apnature$, and that $\apstrength$ must have happened at least once.

The \emph{implicit task} for the player is to learn to read the formulas and to understand how they relate to the trace.
The game offers a help window where the operators are explained.
\autoref{fig:help-screen} shows this help for the \emph{once} modality $\opOnce$, including example formulas and model traces.
But the interface does not \emph{directly} explain the meaning of the formulas on the cards (cf.\ \autoref{subsec:pedagogical-design}).

There are some \emph{nuances} of how the game's logic and $\pltl$ relate.
First, cards in \emph{Tempus fugit} currently do not use \emph{until/since} modalities $\opUntil / \opSince$, although the backend supports them.
Second, the displayed trace is finite, but the game mechanics make it “quasi-infinite” in that cards that globally modify the (finite) future will set default values by which the trace is extended when the trace window grows due to turns arriving at its end.
This makes a “sets $\aplight$ from now on” effect like the one of \emph{Lightning rain} more stable to match players' intuitions.

\subsection{Story}
\label{subsec:story}

The user takes on the role of a young wizard who, accompanied by a fairy, must traverse the \emph{Islands of Tempus Fugit} to cleanse the “corruption” that has plagued the realm's inhabitants.
The protagonist's magic is deeply tied to the flow of time and logic.
Thematically, the wizard needs to \emph{master time}.
$\pltl$ just happens to be the language of magic.

As the protagonist travels from lush forests to scorching deserts and eventually to frozen peaks, they hear whispers of a figure called “The Liberator.”
Levels come with tongue-in-cheek boss battles.
The climax at the “Final Tower” shocks the protagonist by revealing that the Liberator is an older, future version of him.

The game narrative fuels the game in three layers of conflict (cf.~\cite{skolnick2014gameStorytelling} for “conflicts as fuel” in games):
1.~Each level presents a battle against enemies that is resolved by winning the battle.
2.~The battles form an overarching arc to face the “Liberator,” resolved by the final level.
3.~To prevail, protagonist and player have to develop their grasp of magic / temporal logic, which is gradually resolved---although the game's agenda is to just start the players' journeys into the world of logic.

\section{Discussion of design}
\label{sec:design}

\emph{Tempus fugit} is designed to serve as a first contact with formal logic, played for at least 30 to 60~minutes without prior knowledge of the subject.
The challenge of understanding temporal valuations and formulas becomes part of the game.

\subsection{Game design}

\emph{Tempus fugit} is inspired by games like \emph{Slay the Spire}~\cite{SlayTheSpire2019}, a 2019 deck-building roguelike with turn-based combat.
However, except for the draw order, there are no random elements in \emph{Tempus fugit}.
Also, the deck-building aspect is present but not central to the game.

The \emph{core challenge} of the game is how to cleverly combine dealt cards with the possibility of mutating the trace at the current time.
Cards do not interact with each other directly, but only through the trace.
Especially at higher levels, the player has to plan several turns ahead and to strategically manipulate the trace in order to win.

\emph{Fun} arises in various ways:
1.~When the player reaches small subgoals like killing an enemy or making a good card playable.
2.~When the player is lucky enough to draw a card that is applicable in the current situation.
3.~When the player finds a clever way to combine the cards.
4.~When the player enjoys the story humor.

\emph{Player choices} are limited to a clear core: which card to play against which enemy and how to manipulate the current valuation.
When there is no good choice, the player can use their base attack and get an additional card to increase the space of possible actions in the next turn.

\emph{Level progression}
gradually introduces cards and enemies which require the player to understand more and more complex formulas.
Temporal mechanics are introduced step by step.
The first level only uses propositional logic (the $\propPart{\text{red part}}$ of \autoref{def:pltl-syntax}).
The next levels first introduce the one-step-backward modality $\opPrevious$ and then the other past-modalities $\opHistorically, \opOnce$.
Starting on the second island, future-operators feature more and more (the $\opNext, \opGlobally, \opEventually$).
Over time, the nesting of operators becomes more and more complex.
All formulas have been picked to be manageable and such that they can be (non-trivially) satisfied.

\subsection{Pedagogical design and narrative integration}
\label{subsec:pedagogical-design}

The pedagogical design of \emph{Tempus fugit} integrates $\pltl$ syntax into mechanics and narrative.

\subparagraph{Narrative integration of $\pltl$ syntax}
Spells and abilities are expressed through logical formulas, hiding the abstractness of logic in plain sight:
Of course, wizards would use an arcane notation.
Using runes instead of more common symbols deepens this immersion.
In this way, one of the barriers for people approaching formal logic is turned into a feature of the game world and a part of the challenge.
The user is narratively and mechanically motivated to parse complex formulas because they represent the “laws” for successful interaction with the game world.

\subparagraph{Spatio-temporal visualization}

The abstract concept of a trace is represented concretely as a table at the top of the screen.
The constraint that only the current column is directly editable conveys the intuition that the trace unfolds through time.
This helps the user map temporal operators, such as \emph{previously} or \emph{once}, directly onto the sequence of rounds they have played.

\subparagraph*{Semantic scaffolding}

The game introduces different aspects of temporal logic in steps.
Its surrounding mechanics provide a semantic scaffold to avoid overwhelming non-mathematical users with definitions.
The help menu provides more detailed explanations using examples.

\subparagraph*{Seeding logic literacy}

Intentionally, the game does not directly assist the player in reading the formulas (e.g., through natural-language explanations on hover or visualizations of truth conditions on the trace table).
Understanding the formulas is a core part of the game challenge.
Opening the help screen entails friction, which motivates players to memorize what symbols stand for and to find their own intuition.

The game applies only gentle pressure to learn the notation:
Failing to understand a card does not block progress, but understanding more cards is what enables winning at higher levels.
Thus, it is not the case that one loses levels because one fails to parse a specific formula; but one wins levels because one was able to parse sufficiently many formulas!

Understanding $\pltl$ creates game success, even if one does not care about $\pltl$.
But, because the game stays close to the standard notation, the reading ability can be transferred directly to more formal contexts like a course or workshop.
To build firm knowledge of the notation or to transfer to deeper concepts, the embedding in a learning context beyond the game is necessary.

\subsection{Implementation}

\emph{Tempus fugit} can be started directly from \url{https://tempusfugit.equiv.io} and requires less than 10~MB to load.
It runs client-side in the browser and saves all user progress in the browser's local storage.
This makes the game easily accessible in science communication contexts, for instance.
Once loaded, the game does not require an internet connection.

All interactions are handled through mouse clicks or touch input.
However, the screen layout is not optimized for mobile devices, and we recommend playing on a desktop computer or a tablet.

The game has been implemented in TypeScript using the Phaser~3 game framework~\cite{phaser2026repo}, and can readily be modified.
Cards, levels, and enemies are configured through JSON files.
The backend implements a small parser and interpreter for $\pltl$ formulas, which is used to check the applicability of cards and the conditions of enemy attacks.

The internal architecture follows an event-driven design. Since all relevant state changes are initiated by player actions, components communicate exclusively via events rather than direct method calls. In a refined version of this approach, nearly every component can act as an event source. This allows for a loose coupling between game logic and presentation. For instance, when the player loses life points, a corresponding event is emitted, and UI elements such as the life indicator or animations react by subscribing to this event without requiring direct dependencies.

The $\pltl$ interpreter is based on syntax trees. Formulas given in infix notation are translated into postfix form using a modified shunting-yard algorithm and then transformed into abstract syntax trees for evaluation.
% During gameplay, only a finite number of time points are explicitly defined by the player. When evaluating a formula, we therefore restrict computation to these defined states. Evaluation is therefore performed only over these defined states, without introducing an explicit bound or extending the trace. This may lead to undefined behavior at the boundaries of the trace, for instance when temporal operators refer to states beyond the defined range.
% \todo[author=Ben]{I would feel much better if we could communicate what the game does in such situations.}
% While optimizations are possible, like assuming an infinite continuation of false valuations into the future, our use cases did not require such extensions, and the formulas used in the game behaved as intended under this simplified semantics.
% \paragraph{Finite Trace Evaluation and Bounded Horizon.}

In our implementation, formulas are evaluated on a finite sequence of states.
However, linear temporal logic is usually defined over infinite traces.
The formulas in the game are picked in such a way that the nuances of truth values at the end of a finite trace do not play a role.
In particular, the value of a previous operator $\opPrevious$ at the first state is false.

% Only formulas involving the globally operator may be slightly affected by this assumption,
% as they quantify over the entire future.
% All other formulas remain unaffected.
% \todo[inline]{Improve the following sentence}
% Evaluation is therefore restricted to a bounded window together with a sufficiently distant point representing the infinite “false” tail. The required bound is determined by traversing the syntax tree and , which allows us to evaluate temporal operators such as “globally” without constructing an explicit infinite trace.

% Since it is easier for the user to control values in the past than in the future, we extend the operator set by introducing a reverse operator that inverts temporal direction, as well as explicit inverse variants for certain operators (e.g., “next” and “previous”).
%\todo[author=Ben]{The past is already introduced before and maybe does not need to be mentioned here. The reverse op, hopefully, is only an implementation detail? --> yes is only an implementation detail and we can remove this part}
% These additions make formulas easier to interpret for players while remaining close to standard $\pltl$ notation.

% \section{Related work}
%
% .–.\_.–.\_. Would be nice, wouldn't it? .\_.–.\_.–.

\section{Conclusion and related work}
\label{sec:conclusion}

We have presented the design of \emph{Tempus fugit}, a computer game that conveys basic reading skills for temporal logic to unsuspecting people.
It can be used to quickly illustrate how operators of logic interact with valuations and traces.

To our knowledge, \emph{Tempus fugit} is the first computer game to feature temporal logic.
Its use could be combined with more traditional educational games like
\emph{ANDOR}~\cite{jenny2018andor} on expressing oneself in propositional logic, or
\emph{The Incredible Proof Machine}~\cite{breitner2016incrediblePM} on constructing predicate logic proof trees, as well as gamified ones such as
\emph{Turing Complete}~\cite{levelhead2021turingComplete} on logic circuits.
\emph{Tempus fugit} has been a product of a series of courses at TU Berlin where students develop games around formalisms of concurrency theory.
Its sibling projects are available on \url{https://games.equiv.io}.

Gee's idea that games could serve as “learning machines”~\cite{gee2007gamesTeachLearning} has become a classic.
Even in a post-gamified world~\cite{bacalja2024postdigitalVideoGameLiteracy}, a game can be a trick to build engagement.
We feel that there is new relevance for such tricks in a time where people's jobs drift towards specifying systems and evaluating reasoning produced by coding agents.
For this, humans need to grasp formal languages, yet learning such languages becomes less natural as agents take over entry-level tasks.
Framing learning experiences in computer games can address this entry gap, because people understand intuitively that, although a machine could do the task, games are made to be played by humans.

\emph{Tempus fugit} is limited to facilitating a first playful contact.
Activities beyond the game would be necessary to really internalize the syntax and semantics of $\pltl$, and to be able to produce one's own meaningful formulas.
Also, all the intricate topics that a textbook (e.g.~\cite{baierKatoen2008pomc}) or a university course would include are beyond the scope of the game, for instance, Büchi automata, model checking, synthesis, hyperproperties.
But for the first grasp, the game provides an engaging way of getting to know basic concepts.
Although \emph{Tempus fugit} cannot substitute courses on logic, it can generate appetite for such courses.

To contribute to empirical research on educational games~\cite{zeng2020reviewEducationalGames}, it might be worthwhile to study how well people learn using the game.
However, to some extent, the game is its own test:
Anyone who makes it to the end has certainly learned some $\pltl$, and has probably had fun doing so.
We invite everyone to try it out and share their experience with us on \url{https://benkeks.itch.io/tempusfugit}.

\bibliography{bib}

\end{document}